\documentclass{article}
\usepackage{spconf,amsmath,graphicx,hyperref}
\usepackage{multicol}
\usepackage{multirow}
\usepackage{subcaption}

\title{TS-OPD: Reconciling ASR and QA in Speech Language Models via Task-Specific On-Policy Distillation}
\name{Yujie Guo$^{1,\ast}$, Hongjie Chen$^{2}$, Jian Kang$^{2}$, Jie Li$^{2, \dagger}$, Yongxiang Li$^{2}$, Yong Qin$^{1, \dagger}$}
\address{
$^{1}$College of Computer Science, Nankai University, \\
$^{2}$Xingchen AGI Lab, China Telecom Artificial Intelligence Technology Co. Ltd, \\
\small{guoyujie02@mail.nankai.edu.cn, chenhj37@chinatelecom.cn, qinyong@nankai.edu.cn}
}
\begin{document}
\ninept
\maketitle

\begingroup
\renewcommand{\thefootnote}{}
\footnotetext{$^\ast$ This work was done during an internship at Xingchen AGI Lab.}
\footnotetext{$^\dagger$ Yong Qin and Jie Li are corresponding authors.}
\endgroup

%
\begin{abstract}
Speech Language Models (SLMs) inherit strong instruction-following capabilities from pretrained language models, yet ASR specialization can substantially degrade them. To address this ASR--QA trade-off, we propose Task-Specific On-Policy Distillation (TS-OPD), which leverages models before and after ASR specialization as complementary QA and ASR teachers. The student generates separate task-conditioned trajectories for ASR and QA, each supervised only by its corresponding teacher, thereby reducing direct competition between the two supervision signals. Experiments on basic ASR, contextual ASR, and QA demonstrate that TS-OPD improves recognition while preserving QA capability. Moreover, TS-OPD remains robust across different balancing coefficients and continues to benefit from increased distillation data.
\end{abstract}

\section{Introduction}
\label{sec:introduction}

Recent advances in Speech Language Models (SLMs) have extended speech processing beyond conventional ASR toward general speech understanding and interaction~\cite{qwen3omni,stepaudio2.5,salmonn2}.
By connecting a speech encoder with a pretrained LLM, SLMs inherit instruction-following capabilities that can also benefit recognition tasks such as contextual ASR~\cite{contextasr-bench}.
However, ASR-oriented specialization may degrade these capabilities~\cite{qwen3asr}, raising the question of how to improve ASR without sacrificing instruction-following ability.

Existing studies have explored strategies to mitigate capability degradation during adaptation.
Freezing the LLM backbone and speech--text data design can preserve language capabilities during adaptation~\cite{wang2024freeze,alignformer,DeSTA2}, while continual SLM training has explored model merging, reduced LoRA scaling, and experience replay~\cite{hsiao2025analyzing}.
Weight averaging and knowledge distillation have been used to retain capabilities during ASR adaptation~\cite{eeckt2022weight}.
Distillation has been applied to preserve pretrained behaviors and improve speech--text alignment~\cite{blsp-kd,tan2025ssr,wang2026cross}.
These methods preserve prior capabilities through parameter constraints, data design, or distillation.

On-policy distillation (OPD) offers a different perspective.
Originally studied in language model distillation, OPD lets the student generate its own trajectories and queries the teacher on states actually visited by the current student policy, reducing the mismatch between distillation and autoregressive inference~\cite{opd-org,nlpopd1}.
Recent studies have extended OPD to speech models for cross-modal capability transfer and reasoning alignment~\cite{xopd,x3opd}, as well as for retaining recognition capabilities during dialect or multilingual ASR specialization~\cite{opds-mdasr,mopd-mlasr}.
By providing supervision directly on student-visited states, OPD offers a natural mechanism for transferring and preserving complementary behaviors from different model states.

ASR specialization and capability preservation impose competing objectives on SLM adaptation.
We use speech-conditioned QA as a concrete measure of the instruction-following capability retained by the SLM.
Starting from a QA-capable SLM, ASR fine-tuning can improve recognition while degrading QA behavior.
The pre- and post-specialization models therefore provide distinct sources of supervision: the former preserves instruction-following capability, while the latter provides stronger ASR capability.
However, simply combining ASR and QA supervision does not explicitly resolve the interference between these two objectives.
This motivates a task-specific routing mechanism that transfers each capability from its corresponding model state while avoiding direct competition between the two teacher distributions.

In this work, we propose \textbf{Task-Specific On-Policy Distillation (TS-OPD)}, a dual-teacher framework for reconciling ASR and QA capabilities in SLMs.
The original QA-capable model serves as the QA teacher, while its ASR-specialized counterpart serves as the ASR teacher.
The student is initialized from the QA model and generates separate on-policy trajectories for ASR and QA.
Crucially, TS-OPD routes each trajectory only to its corresponding teacher: the ASR teacher transfers recognition capability, while the QA teacher anchors the student to its original QA behavior.
This task-specific routing reduces direct competition between the two teacher distributions and enables the two capabilities to coexist more effectively.
Our main contributions are summarized as follows:

\begin{itemize}
    \item We propose TS-OPD, which uses pre- and post-specialization models as complementary teachers and performs task-specific on-policy distillation to reconcile ASR and QA capabilities.

    \item We characterize the specialization-induced ASR--QA trade-off using a shared-trajectory variant and show that task-specific routing mitigates teacher interference.

    \item Experiments on basic ASR, contextual ASR, and QA demonstrate that TS-OPD improves ASR while preserving QA capability, with favorable robustness and scaling behavior.
\end{itemize}

\begin{figure*}[t]
  \centering
  \includegraphics[width=1.0\textwidth]{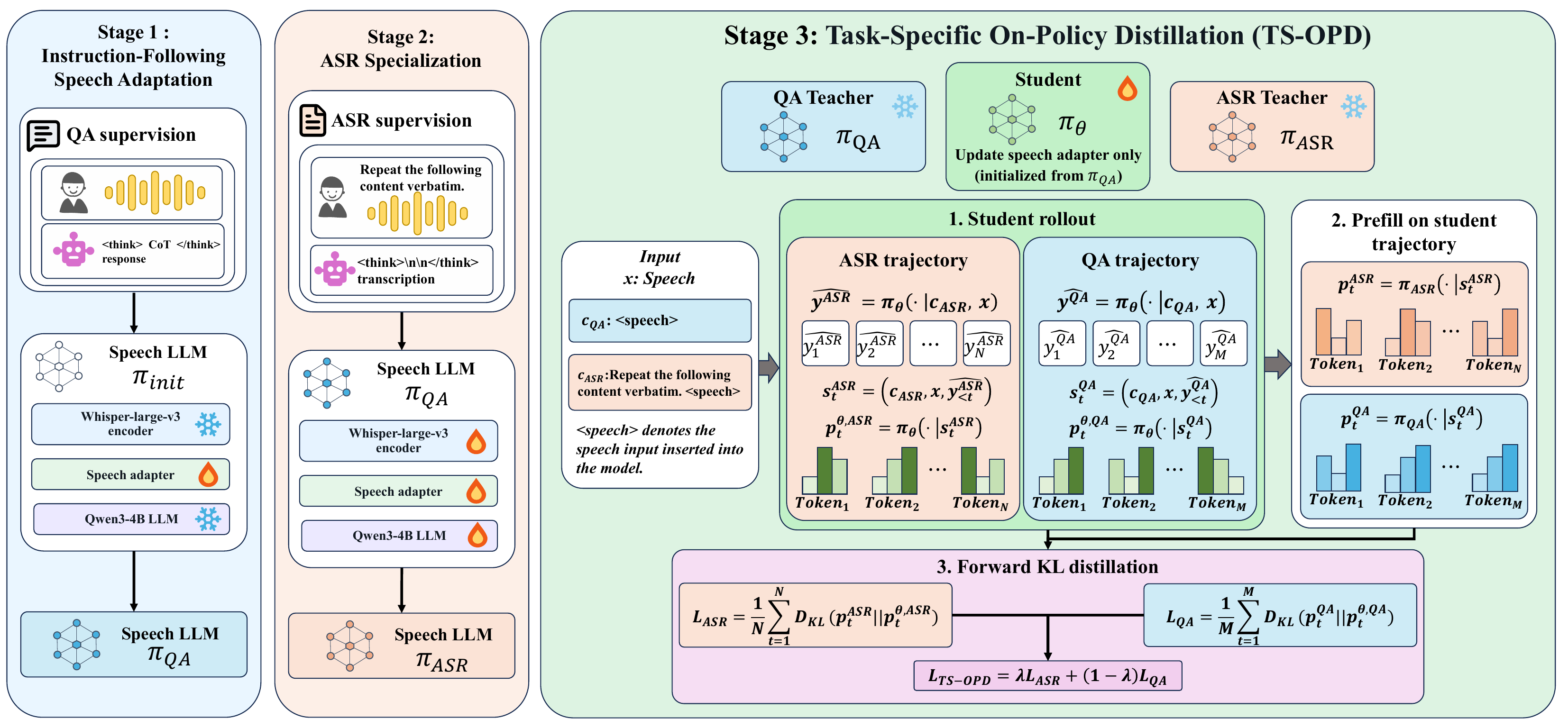}
  \caption{
    Overview of the proposed TS-OPD framework.
    Stage 1 constructs the QA-capable model $\pi_{\mathrm{QA}}$ through instruction-following speech adaptation, while Stage 2 further specializes it for ASR to obtain $\pi_{\mathrm{ASR}}$.
    In Stage 3, the student $\pi_{\theta}$ is initialized from $\pi_{\mathrm{QA}}$ and generates task-specific ASR and QA trajectories.
    Each trajectory is supervised only by its corresponding teacher.
    }
  \label{fig:method}
  \vspace{-3.5mm}
\end{figure*}

\section{Method}
\label{sec:method}

\subsection{Speech LLM Architecture}
\label{sec:speech_llm_arch}

Our Speech Language Model (SLM) follows the widely adopted encoder-adapter-LLM architecture~\cite{qwen3omni,salmonn2}, consisting of a speech encoder, a speech adapter, and a large language model.
We use Whisper-large-v3~\cite{whisper} as the speech encoder and Qwen3-4B~\cite{qwen3} as the LLM.
A lightweight speech adapter composed of convolutional and linear layers connects the two components.
It downsamples the encoder outputs by a factor of four and projects them into the embedding space of the LLM.

\subsection{Teacher and Student Construction}
\label{sec:teacher_student_construction}

Our goal is to improve the ASR capability of an SLM while preserving its QA capability.
As shown in Stage 1 and Stage 2 of Fig.~\ref{fig:method}, we first construct a QA-capable SLM and then further specialize it for ASR.
The resulting models serve as the QA and ASR teachers, respectively, while the student is initialized from the QA model.

In Stage 1, we obtain the QA model $\pi_{\mathrm{QA}}$ by training only the speech adapter with QA supervision while keeping the speech encoder and LLM frozen.
This model serves as the QA teacher.

In Stage 2, starting from $\pi_{\mathrm{QA}}$, we perform full-parameter fine-tuning with ASR supervision to obtain the ASR-specialized model $\pi_{\mathrm{ASR}}$.
This model serves as the ASR teacher.

During TS-OPD, the student $\pi_{\theta}$ is initialized from $\pi_{\mathrm{QA}}$, with both teachers frozen and only its speech adapter updated.

\subsection{Task-Specific On-Policy Distillation}
\label{sec:ts_opd}

We propose Task-Specific On-Policy Distillation (TS-OPD), illustrated in Stage 3 of Fig.~\ref{fig:method}.
The student $\pi_{\theta}$ generates separate on-policy trajectories under the ASR and QA task conditions, and each trajectory is supervised only by its corresponding teacher.

Given a speech $x$, the ASR and QA trajectories are sampled as
\begin{equation}
\begin{aligned}
\hat{\mathbf{y}}^{\mathrm{ASR}}
&=
(\hat{y}^{\mathrm{ASR}}_1,\ldots,\hat{y}^{\mathrm{ASR}}_N)
\sim
\pi_{\theta}(\cdot \mid c_{\mathrm{ASR}},x),\\
\hat{\mathbf{y}}^{\mathrm{QA}}
&=
(\hat{y}^{\mathrm{QA}}_1,\ldots,\hat{y}^{\mathrm{QA}}_M)
\sim
\pi_{\theta}(\cdot \mid c_{\mathrm{QA}},x),
\end{aligned}
\end{equation}

\noindent where $c_{\mathrm{ASR}}$ and $c_{\mathrm{QA}}$ denote the task-specific prompts, and $N$ and $M$ are the ASR and QA response lengths, respectively.

Unlike offline distillation on fixed sequences, TS-OPD performs distillation on the student's own trajectories.
At decoding step $t$, the student-visited states are
\begin{equation}
\begin{aligned}
s_t^{\mathrm{ASR}}
&=
(c_{\mathrm{ASR}},x,\hat{\mathbf{y}}^{\mathrm{ASR}}_{<t}),\quad
s_t^{\mathrm{QA}}
&=
(c_{\mathrm{QA}},x,\hat{\mathbf{y}}^{\mathrm{QA}}_{<t}).
\end{aligned}
\end{equation}

\noindent Under task-specific routing, the ASR trajectory is evaluated only by the ASR teacher, while the QA trajectory is evaluated only by the QA teacher.
The corresponding teacher and student distributions are
\begin{equation}
\begin{aligned}
p_t^{\mathrm{ASR}}
&=
\pi_{\mathrm{ASR}}(\cdot \mid s_t^{\mathrm{ASR}}),
&
p_t^{\theta,\mathrm{ASR}}
&=
\pi_{\theta}(\cdot \mid s_t^{\mathrm{ASR}}),\\
p_t^{\mathrm{QA}}
&=
\pi_{\mathrm{QA}}(\cdot \mid s_t^{\mathrm{QA}}),
&
p_t^{\theta,\mathrm{QA}}
&=
\pi_{\theta}(\cdot \mid s_t^{\mathrm{QA}}).
\end{aligned}
\end{equation}

\noindent We use forward KL divergence for both task-specific objectives:
\begin{equation}
\begin{aligned}
\mathcal{L}_{\mathrm{ASR}}
&=
\frac{1}{N}
\sum_{t=1}^{N}
D_{\mathrm{KL}}
\left(
p_t^{\mathrm{ASR}}
\Vert
p_t^{\theta,\mathrm{ASR}}
\right),\\
\mathcal{L}_{\mathrm{QA}}
&=
\frac{1}{M}
\sum_{t=1}^{M}
D_{\mathrm{KL}}
\left(
p_t^{\mathrm{QA}}
\Vert
p_t^{\theta,\mathrm{QA}}
\right).
\end{aligned}
\end{equation}

\noindent The final TS-OPD objective combines the two task-specific losses:
\begin{equation}
\mathcal{L}_{\mathrm{TS\text{-}OPD}}
=
\lambda \mathcal{L}_{\mathrm{ASR}}
+
(1-\lambda)\mathcal{L}_{\mathrm{QA}},
\end{equation}

\noindent where $\lambda \in [0,1]$ controls their relative contributions.

\begin{table*}[t]
\caption{
Main results on basic ASR (WER), contextual ASR (WER and NE-FNR), and QA (ACC).
A1, A3, and A4 start from randomly initialized adapters. A1 is trained with QA supervision, A3 with ASR only, and A4 jointly with ASR and QA, using each utterance for both tasks. A2 is obtained by full-parameter ASR fine-tuning from A1.
B1 is initialized from A1 and further trained with TS-OPD using 500 hours of distillation data with $\lambda=0.5$.
The best and second-best results are shown in \textbf{bold} and \underline{underlined} text, respectively.
}
\vspace{-1.5mm}
\label{tab:mainres}
\renewcommand{\arraystretch}{1.1}
\centering
\resizebox{\textwidth}{!}{%
\begin{tabular}{c|c|cccc|cc|cccc}
\hline
\multirow{3}{*}{System} & \multirow{3}{*}{Training Strategy} & \multicolumn{4}{c|}{Basic ASR$\downarrow$} & \multicolumn{2}{c|}{ContextASR-Bench} & \multicolumn{4}{c}{QA (TELEVAL)$\uparrow$} \\
 &  & \multirow{2}{*}{\begin{tabular}[c]{@{}c@{}}LS-\\ test-clean\end{tabular}} & \multirow{2}{*}{\begin{tabular}[c]{@{}c@{}}LS-\\ test-other\end{tabular}} & \multirow{2}{*}{GigaSpeech} & \multirow{2}{*}{Overall} & \multirow{2}{*}{WER$\downarrow$} & \multirow{2}{*}{NE-FNR$\downarrow$} & \multirow{2}{*}{LlamaQA-en} & \multirow{2}{*}{TriviaQA-en} & \multirow{2}{*}{WebQ-en} & \multirow{2}{*}{Overall} \\
 &  &  &  &  &  &  &  &  &  &  &  \\ \hline
A1 & QA Adapter & 5.26 & 10.15 & 15.91 & 14.19 & 9.38 & {\underline{12.28}} & 71.67 & 34.77 & 41.43 & 42.57 \\
A2 & A1 ASR Full-FT & \textbf{2.59} & \textbf{5.50} & \textbf{10.52} & \textbf{9.16} & \textbf{5.65} & 20.31 & 1.67 & 4.78 & 2.22 & 2.86 \\
A3 & ASR Adapter & 4.02 & 8.03 & 13.04 & 11.57 & 9.29 & 26.02 & 0.33 & 2.63 & 1.81 & 1.89 \\
A4 & ASR+QA Adapter & 5.09 & 9.52 & 15.08 & 13.44 & 8.70 & \textbf{10.54} & 73.33 & \textbf{36.68} & {\underline{43.09}} & {\underline{44.29}} \\
B1 & TS-OPD & {\underline{3.55}} & {\underline{7.15}} & {\underline{11.28}} & {\underline{10.03}} & {\underline{6.19}} & 16.82 & \textbf{74.00} & {\underline{35.48}} & \textbf{44.74} & \textbf{45.07} \\ \hline
\end{tabular}
}
\vspace{-3.5mm}
\end{table*}

\begin{table}[t]
\caption{
Ablation of teacher supervision and routing strategies.
B3 uses only the ASR teacher, while B2 and B1 use both teachers.
TS and ST denote TS-OPD and ST-OPD, defined in Secs.~\ref{sec:ts_opd} and~\ref{sec:ablation}, respectively.
The best and second-best results are shown in \textbf{bold} and \underline{underlined} text, respectively.
}
\vspace{-1.5mm}
\label{tab:ablation}
\renewcommand{\arraystretch}{1.1}
\centering
\resizebox{0.49\textwidth}{!}
{%
\begin{tabular}{c|ccc|cccc}
\hline
\multirow{3}{*}{System} & \multicolumn{3}{c|}{Training Strategy} & ASR & \multicolumn{2}{c}{ContextASR-Bench} & QA \\
 & Teacher & Teacher & \multirow{2}{*}{Routing} & \multirow{2}{*}{Overall$\downarrow$} & \multirow{2}{*}{WER$\downarrow$} & \multirow{2}{*}{NE-FNR$\downarrow$} & \multirow{2}{*}{Overall$\uparrow$} \\
 & ASR & QA &  &  &  &  &  \\ \hline
A1 & - & - & - & 14.19 & 9.38 & \textbf{12.28} & 42.57 \\
B3 & $\surd$ & - & - & \textbf{9.96} & \underline{6.84} & 22.37 & 1.89 \\
B2 & $\surd$ & $\surd$ & ST & 10.92 & 8.96 & 17.38 & \underline{42.73} \\
B1 & $\surd$ & $\surd$ & TS & \underline{10.03} & \textbf{6.19} & \underline{16.82} & \textbf{45.07} \\ \hline
\end{tabular}
}
\vspace{-3.5mm}
\end{table}

\section{Experiments}
\label{sec:experiments}

\subsection{Datasets and Experimental Setup}
\label{sec:datasets_setup}

\textbf{Training data.}
We sample approximately 8k hours of English speech from Emilia~\cite{emilia}, paired with ASR transcriptions.
We use each transcription as a textual query to Qwen3-4B~\cite{qwen3} and take its chain-of-thought response as QA supervision, yielding paired ASR and QA targets for the same speech input.

\textbf{Prompt and response format.}
As shown in Stages 1 and 2 of Fig.~\ref{fig:method}, we use task-specific prompts for ASR and QA.
ASR uses a verbatim repetition prompt with an empty reasoning span followed by the transcription, while QA directly takes the speech input and generates chain-of-thought reasoning followed by the final response.

\textbf{Evaluation.}
For basic ASR evaluation, we use LibriSpeech (LS) test-clean and test-other~\cite{librispeech} together with the GigaSpeech test set, and report word error rate (WER (\%)).
For contextual ASR, we use the English Dialogue subset of ContextASR-Bench~\cite{contextasr-bench}. The corresponding hotword list is appended to the standard ASR prompt as additional context, and we report WER (\%) and NE-FNR (\%).
For QA evaluation, we use LlamaQA-en, TriviaQA-en, and WebQ-en from TELEVAL~\cite{televal}, and report accuracy (ACC (\%)).
Long-form dialogue recordings in ContextASR-Bench are segmented into shorter utterances, and samples longer than 30 seconds are excluded to satisfy the input-length constraint of the speech encoder.

\textbf{Experimental setup.}
We construct four SLM variants.
A1 is trained on 8k-hour QA data by updating only the speech adapter, and A2 is obtained by full-parameter ASR fine-tuning from A1.
A3 and A4 use randomly initialized adapters with the encoder and LLM frozen; A3 is trained with ASR only, while A4 uses paired ASR and QA supervision for each utterance.

For TS-OPD, A1 and A2 serve as the frozen QA and ASR teachers, respectively, and the student is initialized from A1 with only its speech adapter updated.
ASR and QA instances are sampled at a 1:1 ratio and routed to their corresponding teachers.
We use top-128 forward KL and train for one epoch with a learning rate of $1\times10^{-5}$ and a cosine scheduler.
The maximum generation lengths are 256 and 2048 tokens for ASR and QA, respectively.

\subsection{Main Results}
\label{sec:main_results}

Table~\ref{tab:mainres} reports the main results. For TS-OPD, we use 500 hours sampled from the Emilia training set with $\lambda=0.5$.

\textbf{TS-OPD improves basic ASR while preserving QA capability.}
A1 and A4 achieve strong QA performance but have relatively weak ASR, whereas A2 achieves the best ASR through full-parameter specialization but nearly loses QA capability.
In contrast, TS-OPD (B1) narrows the ASR gap to A2 and consistently outperforms the ASR-only adapter baseline A3, while retaining strong QA capability.
Notably, A4 uses joint ASR and QA supervision over the full 8k-hour corpus, yet remains substantially weaker than B1, which uses only 500 hours during the TS-OPD stage.
This suggests that simple balanced multi-task training is insufficient to reconcile the two capabilities, whereas TS-OPD more effectively transfers ASR capability while preserving QA behavior.

\textbf{TS-OPD achieves a better balance on contextual ASR.}
A2 obtains the lowest WER, but its relatively high NE-FNR indicates weaker named-entity recall.
Conversely, A1 and A4 achieve lower NE-FNR but considerably higher WER.
B1 provides a more balanced result between transcription accuracy and named-entity recognition, suggesting that the preserved language capability remains useful for contextual speech recognition.

\textbf{Improved ASR does not come at the expense of QA capability.}
B1 not only preserves QA performance, but also achieves the highest overall TELEVAL score among all systems.
This indicates that TS-OPD can improve speech recognition without degrading speech-conditioned QA performance, and may even slightly enhance it.
With only 500 hours of distillation data, TS-OPD effectively refines the speech adapter toward stronger ASR capability while preserving its QA capability.

\begin{figure}[t]
\centering

\begin{subfigure}{0.8\linewidth}
    \centering
    \includegraphics[width=\linewidth]{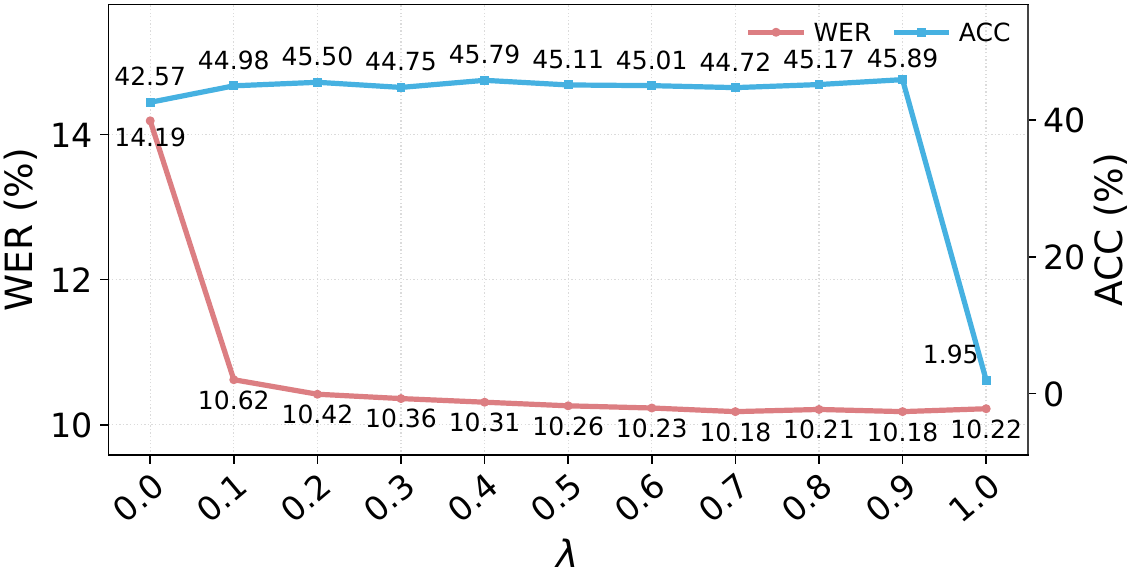}
    \caption{TS-OPD with 300 hours of distillation data.}
    \label{fig:ts-lambda}
\end{subfigure}

\vspace{0.5em}

\begin{subfigure}{0.8\linewidth}
    \centering
    \includegraphics[width=\linewidth]{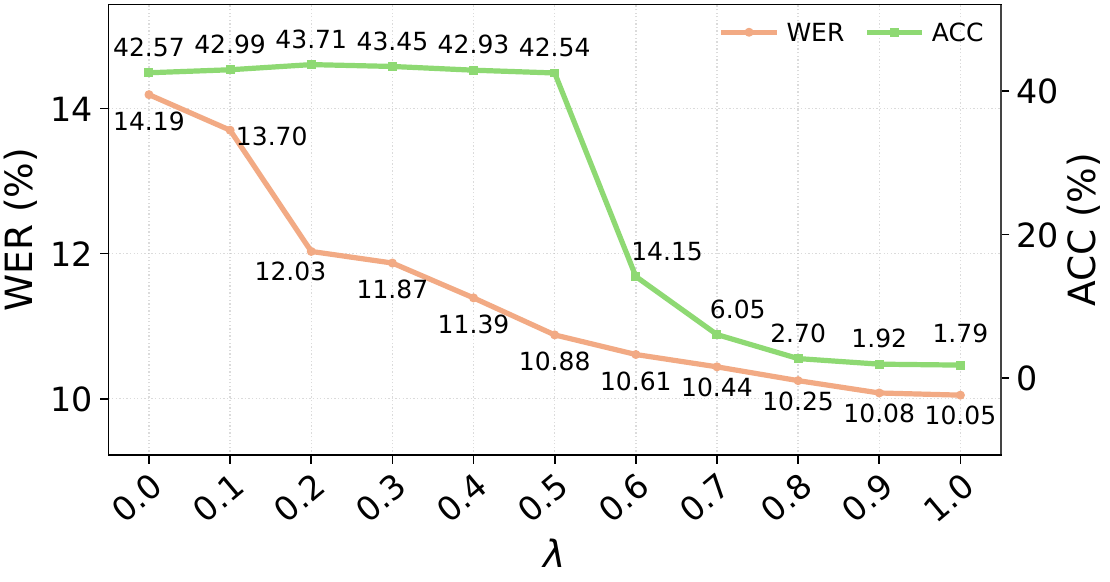}
    \caption{ST-OPD with 500 hours of distillation data.}
    \label{fig:st-lambda}
\end{subfigure}

\caption{
Effect of $\lambda$ on TS-OPD and ST-OPD.
TS-OPD and ST-OPD use 300 and 500 hours of Emilia speech, respectively.
The comparison focuses on their performance trends with respect to $\lambda$.
}
\vspace{-3mm}
\label{fig:lambda}

\end{figure}
\subsection{Ablation Study}
\label{sec:ablation}

Table~\ref{tab:ablation} studies the effects of teacher supervision and routing strategy.
All distillation variants are initialized from A1 and use the same 500-hour Emilia subset and optimization settings.

\textbf{Teacher supervision and routing.}
B3 uses only the ASR teacher without QA supervision.
For the dual-teacher variants, we construct Shared-Trajectory OPD (ST-OPD) as a controlled variant to analyze the interaction between the two teacher distributions, and compare it with the proposed Task-Specific OPD (TS-OPD).

In ST-OPD (B2), every student trajectory is evaluated by both teachers, regardless of its task identity.
Given a trajectory under task condition $z\in\{\mathrm{ASR},\mathrm{QA}\}$, the ASR and QA teachers evaluate the same student-visited state $s_t^z$:

\begin{equation}
\begin{aligned}
p_{t,z}^{\mathrm{ASR}}
&=
\pi_{\mathrm{ASR}}(\cdot\mid s_t^z), \quad
p_{t,z}^{\mathrm{QA}}
&=
\pi_{\mathrm{QA}}(\cdot\mid s_t^z).
\end{aligned}
\end{equation}

\noindent Both teacher distributions supervise the same student distribution via

\begin{equation}
\begin{aligned}
\mathcal{L}_{\text{ST-OPD}}
={}&
\frac{1}{T_z}
\sum_{t=1}^{T_z}
\Bigg[
\lambda
D_{\mathrm{KL}}
\left(
p_{t,z}^{\mathrm{ASR}}
\Vert
p_t^{\theta,z}
\right)
\\
&\qquad+
(1-\lambda)
D_{\mathrm{KL}}
\left(
p_{t,z}^{\mathrm{QA}}
\Vert
p_t^{\theta,z}
\right)
\Bigg],
\end{aligned}
\end{equation}

\noindent where $T_z=N$ for ASR and $T_z=M$ for QA.

In contrast, TS-OPD (B1) routes each trajectory according to its task identity:
ASR trajectories are evaluated only by $\pi_{\mathrm{ASR}}$, while QA trajectories are evaluated only by $\pi_{\mathrm{QA}}$, as defined in Sec.~\ref{sec:ts_opd}.
Thus, ST-OPD applies both teacher distributions to every trajectory, whereas TS-OPD uses task-specific teacher routing.

\textbf{The QA teacher anchors capability during ASR transfer.}
B3, distilled only from the ASR teacher, achieves strong ASR performance but nearly loses its QA capability.
Since the student is initialized from A1, the ASR-only objective continuously shifts it toward the ASR-specialized teacher A2.
In TS-OPD, the QA teacher anchors the student to its original QA behavior while the ASR teacher transfers recognition capability.
As a result, B1 achieves nearly the same basic ASR performance as B3 while preserving QA capability.

\textbf{Task-specific routing reduces interference between the two teachers.}
Compared with ST-OPD, TS-OPD performs better on basic ASR, contextual ASR, and QA.
In ST-OPD, both teachers supervise the same student trajectory, so their different output distributions may provide conflicting signals at the same state.
TS-OPD instead routes ASR and QA trajectories to their corresponding teachers, reducing direct interference between the two supervision signals.
This enables more effective coexistence of ASR and QA capabilities.

Compared with A1, TS-OPD substantially improves ASR while slightly improving QA performance.
These results highlight the importance of complementary teacher supervision and task-specific routing for reconciling the two capabilities.

\subsection{Effect of $\lambda$ under Different Routing Strategies}

Fig.~\ref{fig:lambda} compares the effect of $\lambda$ on TS-OPD and ST-OPD.
The two experiments use 300 and 500 hours of distillation data, respectively, so our primary focus is on their sensitivity to $\lambda$ rather than a strictly controlled absolute comparison.
Nevertheless, TS-OPD with only 300 hours already achieves a more favorable ASR–QA trade-off than ST-OPD with 500 hours.

\textbf{TS-OPD is robust to the choice of $\lambda$.}
For $\lambda$ between 0.1 and 0.9, both ASR and QA remain stable, with QA accuracy varying by only about one percentage point.
This behavior is consistent with task-specific routing: because the two teachers supervise different task-conditioned trajectories, $\lambda$ mainly adjusts the relative strengths of the two task objectives rather than directly balancing two teacher distributions at the same student-visited state.
Only when $\lambda=1$, where QA supervision is removed, does QA capability collapse.

\textbf{ST-OPD reveals an explicit ASR--QA trade-off.}
In ST-OPD, both teachers supervise the same student trajectory, so $\lambda$ directly controls their relative influence on the shared student distribution.
As $\lambda$ increases, ASR improves while QA eventually degrades sharply.
Among the tested values, $\lambda=0.5$ yields the most favorable empirical trade-off between ASR and QA, corresponding to symmetric weighting of the two teachers.
Interestingly, increasing $\lambda$ to $0.55$ already reduces QA accuracy to $23.97\%$, with an ASR WER of $10.82\%$.
This sensitivity is consistent with competition between the two teacher distributions under shared-trajectory supervision.

Overall, ST-OPD explicitly trades off ASR acquisition and QA preservation through $\lambda$, whereas TS-OPD separates the two supervision signals by task and is therefore much less sensitive to the balancing coefficient.

\subsection{Scaling with Distillation Data}

\begin{figure}[t]
\centering

\begin{subfigure}[t]{0.48\linewidth}
    \centering
    \includegraphics[width=\linewidth]{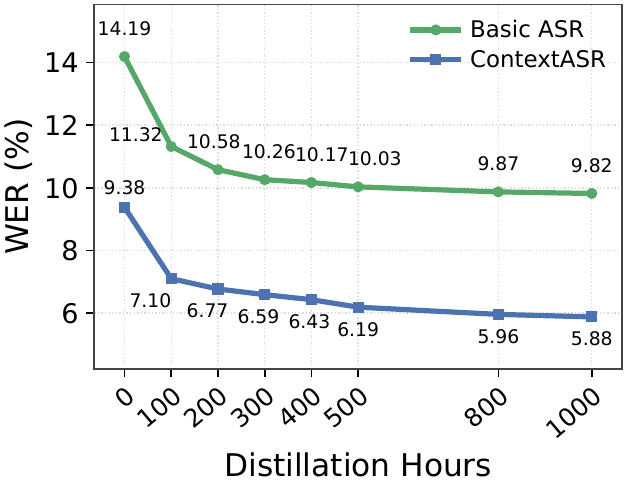}
    \caption{ASR performance.}
    \label{fig:scale-asr}
\end{subfigure}
\hfill
\begin{subfigure}[t]{0.48\linewidth}
    \centering
    \includegraphics[width=\linewidth]{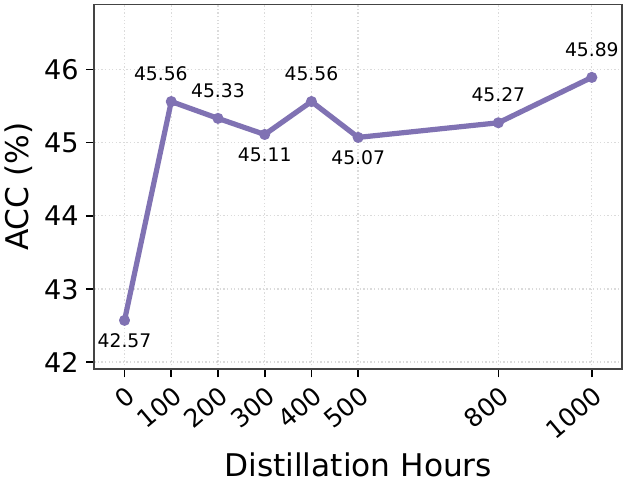}
    \caption{QA performance.}
    \label{fig:scale-qa}
\end{subfigure}

\caption{
Scaling behavior of TS-OPD with different amounts of distillation data.
}

\vspace{-3mm}
\label{fig:scale}

\end{figure}

Fig.~\ref{fig:scale} studies how TS-OPD scales with the amount of distillation data.
As the data increases from 100 to 1000 hours, both basic ASR and contextual ASR improve consistently.
The overall basic ASR WER decreases from $11.32\%$ to $9.82\%$, while the contextual ASR WER decreases from $7.10\%$ to $5.88\%$.
In contrast, QA accuracy remains highly stable throughout scaling, staying around $45\%$.

These results show that increasing the distillation data steadily strengthens recognition capability without degrading QA performance.
Notably, substantial ASR gains are already obtained with relatively little data, while further scaling continues to provide consistent improvements.

\section{Conclusion}

In this work, we propose TS-OPD to reconcile ASR and QA capabilities by using complementary pre- and post-specialization teachers with task-specific trajectory routing.
Experiments show that TS-OPD improves basic and contextual ASR while preserving QA capability, mitigates the ASR--QA trade-off observed in shared-trajectory OPD, and remains robust to $\lambda$ with favorable scaling behavior.




\bibliographystyle{IEEEbib}
\bibliography{strings,refs}

\end{document}